\documentclass[a4paper,10pt,twoside]{cpc-hepnp}

\usepackage{multicol}
\usepackage{graphicx}
\usepackage{amssymb,bm,mathrsfs,bbm,amscd}
\usepackage[tbtags]{amsmath}
\usepackage{lastpage}
\usepackage{epsfig}

\long\def\inst#1{\par\nobreak\kern 4pt\nobreak
    {\itshape #1}\par\vskip 10pt plus 3pt minus 3pt}
\RequirePackage{xspace}

\def\babar{\mbox{\slshape B\kern-0.1em{\smaller A}\kern-0.1em
    B\kern-0.1em{\smaller A\kern-0.2em R}}}
\def\Kbar    {\kern 0.18em\overline{\kern -0.18em K}{}\xspace}

\def\Kz      {\ensuremath{K^0}\xspace}
\def\Kzb     {\ensuremath{\Kbar^0}\xspace}
\def\KzKzb   {\ensuremath{\Kz {\kern -0.16em \Kzb}}\xspace}

\def\Ks     {\ensuremath{K_S}\xspace}
\def\Kl     {\ensuremath{K_L}\xspace}
\def\KsKs   {\ensuremath{\Ks {\kern -0.16em \Ks}}\xspace}
\def\KlKl   {\ensuremath{\Kl {\kern -0.16em \Kl}}\xspace}
\def\KsKl   {\ensuremath{\Ks {\kern -0.16em \Kl}}\xspace}
\def\KlKs   {\ensuremath{\Kl {\kern -0.16em \Ks}}\xspace}
\def\Dbar    {\kern 0.18em\overline{\kern -0.18em D}{}\xspace}
\def\Dz      {\ensuremath{D^0}\xspace}
\def\Dzb     {\ensuremath{\Dbar^0}\xspace}
\def\DzDzb   {\ensuremath{\Dz {\kern -0.16em \Dzb}}\xspace}
\newcommand{\DD}{\ensuremath{D\Dbar}\xspace}
\newcommand{\DsP}{\ensuremath{D_s^+}\xspace}
\newcommand{\DsM}{\ensuremath{D_s^-}\xspace}
\newcommand{\DspDsm}{\ensuremath{\DsP {\kern -0.16em \DsM}}\xspace}
\newcommand{\Dp}{\ensuremath{D^+}\xspace}
\newcommand{\Dm}{\ensuremath{D^-}\xspace}
\newcommand{\DpDm}{\ensuremath{\Dp {\kern -0.16em \Dm}}\xspace}
\def\Bbar    {\kern 0.18em\overline{\kern -0.18em B}{}\xspace}

\def\Bz      {\ensuremath{B^0}\xspace}
\def\Bzb     {\ensuremath{\Bbar^0}\xspace}
\def\BzBzb   {\ensuremath{\Bz {\kern -0.16em \Bzb}}\xspace}
\def\Bu      {\ensuremath{B^+}\xspace}
\def\Bub     {\ensuremath{B^-}\xspace}

\def\BpBm    {\ensuremath{\Bu {\kern -0.16em \Bub}}\xspace}

\def\Ds      {\ensuremath{D^+_s}\xspace}
\def\Dp      {\ensuremath{D^+}\xspace}

\newcommand{\optbar}[1]{\shortstack{{\tiny (\rule[.4ex]{1em}{.1mm})}
  \\ [-.7ex] $#1$}}
\def\BorBbar    {\kern 0.18em\optbar{\kern -0.18em B}{}\xspace}
\def\DorDbar    {\kern 0.18em\optbar{\kern -0.18em D}{}\xspace}
\def\KorKbar    {\kern 0.18em\optbar{\kern -0.18em K}{}\xspace}

\def\pep2{PEP-II}
\mathchardef\Upsilon="7107
\def\Y#1S{\ensuremath{\Upsilon{(#1S)}}\xspace}

\begin{document}

\title{A possible signature of new physics at BES-III
\thanks{Supported by National Natural Science Foundation of China
£¨10575108,10521003, 10735080£©, the 100 Talents program of CAS,
and the Knowledge Innovation Project of CAS under contract Nos.
U-612 and U-530 (IHEP)} }

\author{%
ZOU Jia-Heng$^{1)}$\email{zoujh@ihep.ac.cn} \quad  LI
Hai-Bo$^{2)}$\email{lihb@ihep.ac.cn} \quad ZHANG
Xue-Yao$^{1)}$\email{zhangxy@hepg.sdu.edu.cn} }

\maketitle

\address{1~(Physics Department, Shan Dong University, Jinan,
250100, China)\\
2~(Institute of High Energy Physics, P.O.Box 918, Beijing 100049,
China)}

\begin{abstract}
The recent observations of the purely leptonic decay $\Ds
\rightarrow \mu^+ \nu_{\mu}$ and $\tau^+\nu_{\tau}$ at CLEO-c and
$B$ factory may allow a possible contribution from a charged Higgs
boson. One such measurement of the decay constant $f_{D_s}$
differs from the most precise unquenched lattice QCD calculation
by a level of 4 $\sigma$. Meanwhile, the measured ratio, ${\cal
BR}(\Ds \rightarrow \mu^+ \nu_{\mu})$/${\cal BR}(\Dp \rightarrow
\mu^+ \nu_{\mu})$, is larger than the standard model prediction at
a 2.0$\sigma$ level.  We discuss that the precise measurement of
the ratio ${\cal BR}(\Ds \rightarrow \mu^+ \nu_{\mu})$/${\cal
BR}(\Dp \rightarrow \mu^+ \nu_{\mu})$ at BES-III will shed light
on the presence of new intermediate particles by comparing the
data with the theoretical predictions, especially, the predictions
of high precise unquenched lattice QCD calculations.
\end{abstract}

\begin{keyword}
BES-III, decay constant, QCD, leptonic decay
\end{keyword}

\begin{pacs}
13.30.Ce, 13.20.Fc
\end{pacs}

\footnotetext[0]{\hspace*{-2em}\small\centerline{\thepage\ --- \pageref{LastPage}}}%

\begin{multicols}{2}

\section{Introduction}

Purely leptonic decays of heavy mesons are of great interest both
theoretically and experimentally. Measurements of the decays $B^+
\rightarrow l^+ \nu$, $\Ds \rightarrow l^+\nu$ and $\Dp
\rightarrow l^+\nu$,  provide an experimental determination of the
product of CKM elements and decay constants.  If the CKM element
is measured from other reactions, the leptonic decays can access
the decay constants, which can be used to test lattice QCD
predictions for heavy quark systems.

In the Standard Model (SM) the purely leptonic decays $B^+
\rightarrow  l^+ \nu$ and $\Ds \rightarrow l^+ \nu$ proceed via
annihilation of the heavy meson into a $W^*$. Akeroyd
 and Chen~\cite{akeroyd2007} pointed out that the leptonic decay widths are
 modified by new physics. For the $D^+$ and $\Ds$,
 the two $SU(2)_L \times U(1)_Y$ Higgs doublets with
 hypercharge $Y=1$ (2HDM) would contribute to these decays~\cite{akeroyd2007}.
 The tree level partial width in the 2HDM is given by~\cite{akeroyd2007}
\begin{eqnarray}
 \Gamma(\Ds \rightarrow l^+ \nu) &= & \frac{G^2_F m_{\Ds} m^2_l
 f^2_{\Ds}}{8\pi} |V_{cs}|^2 \left(1-\frac{m_l^2}{m^2_{\Ds}}
 \right)^2 \nonumber \\
& &\times r_s,
 \label{eq:lepton-amp}
\end{eqnarray}
where $G_F= 1.16639 \times 10^{-5} \, \mbox{GeV}^{-2}$ is the
Fermi constant, $m_l$ is the mass of the lepton, $m_{\Ds}$ is the
mass of the $\Ds$ meson, $V_{cs}$ is the Cabibbo-Kobayashi-Maskawa
(CKM) matrix element, and $f_{\Ds}$ is the decay constant. In the
2HDM (model II type Yukawa couplings), at tree level, the scaling
factor $r_s$ is given by~\cite{akeroyd2007}
\begin{eqnarray}
 r_s &=& \left[ 1- m^2_{\Ds} \frac{\mbox{tan}^2\beta}{m^2_{H^{\pm}}}
 \left(\frac{m_s}{m_c+m_s}\right) \right]^2 \nonumber \\
 &=& \left[ 1- m^2_{\Ds}R^2 \left(\frac{m_s}{m_c+m_s}\right)
 \right]^2,
 \label{eq:define}
\end{eqnarray}
where $m_{H^\pm}$ is the charged Higgs mass, $m_c$ is the charm
quark mass, $m_s$ is the strange quark mass (for $\Dp$ decays, it
is the light $d$-quark mass), $\mbox{tan}\beta$ is the ratio of
the vacuum expectation values of the two Higgs doublets,  and the
$H^\pm$ contribution to the decay rate depends on $R =
\frac{\mbox{tan}\beta}{m_{H^{\pm}}}$. The contribution from the
$H^{\pm}$ interferes destructively with the $W^{\pm}$ mediated SM
diagram. As discussed in reference~\cite{rosner2008}, the recent
experimental measurements of ${\cal BR}(B^\pm \rightarrow \tau^\pm
\nu_{\tau})$~\cite{belleB2006,babar2007} provide an upper limit of
$R<0.29$ GeV$^{-1}$ at 90\% C.L.. For values of $R$ in the
interval $0.20<R<0.30$ GeV$^{-1}$,  the charged Higgs contribution
could have a sizable effect on  the $\Ds$ leptonic decay
rate~\cite{akeroyd2007,rosner2008}.
For the quark masses $m_s$ and $m_c$ the range of $0.03 <
m_s/(m_c+m_s)< 0.15$ is used in the following discussions based on
the Particle Data Group values~\cite{akeroyd2007,pdg2007}.

For the $\Dp$, $m_d << m_c$, the modification is negligible, and
thus the scaling factor $r_d \approx 1$. However, in the case of
the $\Ds$, the scaling factor $r_s$ may be sizable due to the
non-negligible $m_s /m_c$.  Although the contribution of the new
physics to the rate is small in comparison to the SM rate for $\Ds
\rightarrow l^+\nu$ decays,  measureable effects may be accessible
since the decay rate for $\Ds \rightarrow \mu^+\nu_{\mu}$ is much
larger than that for $B$ leptonic decays, and can be measured with
good precision.

\section{Recent measurements and constraints }
The most precise measurement of the branching fraction for the
$\Dp \rightarrow \mu \nu_{\mu}$ is from CLEO-c based on
281$pb^{-1}$ of data taken on the $\psi(3770)$ peak. The measured
decay rate of the $\Dp \rightarrow \mu \nu_{\mu}$ is $(4.40 \pm
0.66^{+0.09}_{-0.12})\times 10^{-4}$~\cite{cleocfd}.  In the
context of the SM, using the well measured $\Dp$ lifetime of
$1.040 \pm 0.007$ ps and assuming $|V_{cd}| =|V_{us}| =
0.2238(29)$, they determine~\cite{cleocfd}
\begin{eqnarray}
(f_{\Dp})_{\mbox{CLEO-c}} = (222.6 \pm 16.7^{+2.8}_{-3.4})\,\,
\mbox{MeV}.
 \label{eq:dp}
\end{eqnarray}

Recently,  measurements of $\Ds \rightarrow l^+\nu$ decays with
precision levels comparable to that for $\Dp \rightarrow \mu^+\nu$
decays have been reported by CLEO-c~\cite{cleoc2007,cleoc2008},
BaBar~\cite{babar2007d} and Belle~\cite{belle2007,rosner2008}. For
the $\Ds \rightarrow \mu \nu_{\mu}$ decay mode, the combined decay
rate from the CLEO-c, Belle and BaBar experiments is
$(6.26\pm0.43\pm0.25)\times 10^{-3}$. For the $\Ds \rightarrow
\tau^+ \nu_{\tau}$ decay mode, combining the two $\tau$ decay
channels ($\tau^+ \rightarrow \pi^- \bar{\nu}_{\tau}$ and
$e^+\nu_{e} \bar{\nu}_{\tau}$) from CLEO-c~\cite{cleoc2008}, one
obtains ${\cal B}(\Ds \rightarrow \tau^+
\nu_{\tau})=(6.47\pm0.61\pm 0.26)\%$. Using the $\Ds$ lifetime of
0.50 ps and $|V_{cs}| =0.9737$~\cite{pdg2007} in the SM relation,
one determines the decay constant $f_{\Ds}$ from the $\Ds
\rightarrow \mu^+ \nu_{\mu}$ mode to be
\begin{eqnarray}
(f_{\Ds})^{\mu}_{exp} = (272 \pm 11) \, \mbox{MeV},
 \label{eq:fds_exp_mu}
\end{eqnarray}
and that from $\Ds \rightarrow \tau^+ \nu_{\tau}$ decay mode to be
\begin{eqnarray}
(f_{\Ds})^{\tau}_{exp} = (285 \pm 15) \, \mbox{MeV}.
 \label{eq:fds_exp_tau}
\end{eqnarray}
The average of $\tau \nu_{\tau}$ and $\mu \nu_{\mu}$ values is
\begin{eqnarray}
(f_{\Ds})_{exp} = (276 \pm 9)\, \mbox{MeV}.
 \label{eq:fds_exp_ave}
\end{eqnarray}
Recently, the HPQCD+UKQCD collaboration claims better than 2\%
precision for their unquenched calculations~\cite{hpqcd2008}
\begin{eqnarray}
(f_{\Dp})_{QCD} &= &(208 \pm 4) \mbox{MeV}, \nonumber \\
(f_{\Ds})_{QCD} &= &(241\pm 3)\mbox{MeV},
 \label{eq:fds_theo}
\end{eqnarray}
which is four times better than the experiment and previous
theory~\cite{fnal2005,qcdsf,taiwan2005,ukqcd2001}. As pointed out
in Ref.~\cite{bogdan2008}, there is a 15\% ($3.8\sigma$)
discrepancy between the experimental and lattice QCD values of
$f_{\Ds}$ (Eqs. (\ref{eq:fds_exp_ave}) and (\ref{eq:fds_theo})).
The discrepancy is seen in both the $\tau\nu_{\tau}$ mode, where
it is 18\% (2.9 $\sigma$), and the  $\mu\nu_{\mu}$ where it is
13\% (2.7 $\sigma$).

Equation (\ref{eq:lepton-amp}) shows that the charged Higgs would
lower the $\Ds$ decay rate relative to the SM prediction. However,
the LQCD predicted value (Eq. (\ref{eq:fds_theo})) is {\em below}
the measured value by more than 3$\sigma$.   This indicates that
there is no value of $m_{H^+}$ in the 2HDM that can accommodate
the measured $f_{D_s}$ value~\cite{rosner2008}.  If we take the
discrepancy seriously, there must be new physics that  {\em
enhances} the predicted leptonic decay rate.

Measurements of $f_{\Ds}$ (a detailed summary in
Ref.~\cite{rosner2008}) and its world average are shown in
Fig.~\ref{fig:exp:fds} together with the LQCD prediction. With 20
fb$^{-1}$ at $E_{CM} = 4170$ MeV, the BES-III sensitivity for the
measurement of the leptonic $\Ds$ decay branching fraction would
be about 2\%~\cite{li2006}, which corresponds to a 1.0\%
uncertainty level for $f_{\Ds}$, as indicated in
Fig.~\ref{fig:exp:fds}.  Assuming that the central value for the
combined experimental $f_{\Ds}$ result persists, the discrepancy
between the SM prediction and a BES-III measurement would be more
than 8$\sigma$, and a signal for new physics beyond the SM.

\begin{center}
\includegraphics[scale=0.4]{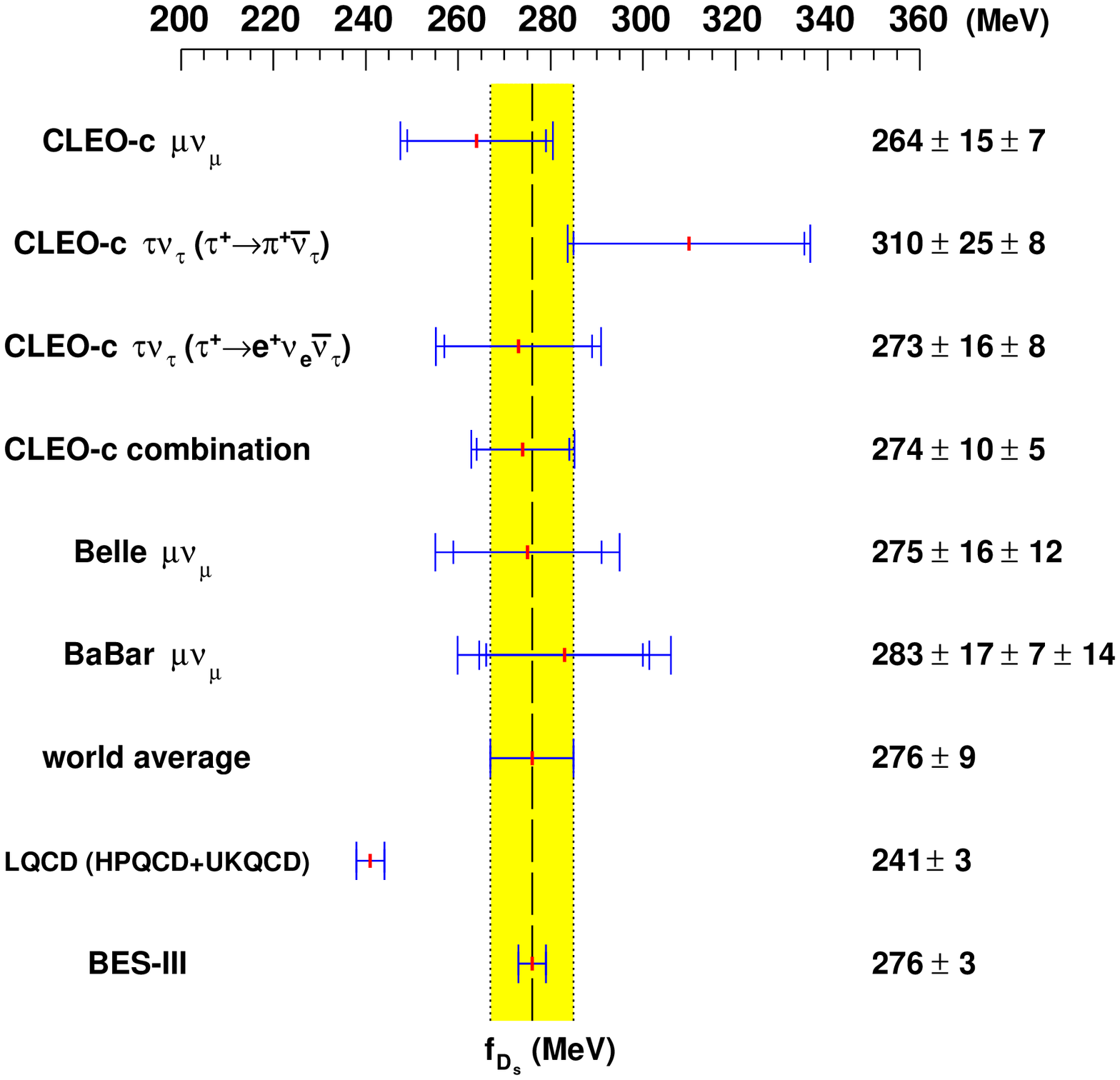}
\figcaption{\label{fig:exp:fds}Values of $f_{\Ds}$ extracted from
different experiments in the context of the SM (a detailed summary
in Ref.~\citep{rosner2008}). The world average is $f_{\Ds} = 276
\pm 9$ MeV, with an uncertainty of about 3.3\%.
 The 1\% BES-III sensitivity to $f_{\Ds}$   is
indicated with the assumption that the current world average
central value persists.}
\end{center}
\begin{center}
\includegraphics[scale=0.4]{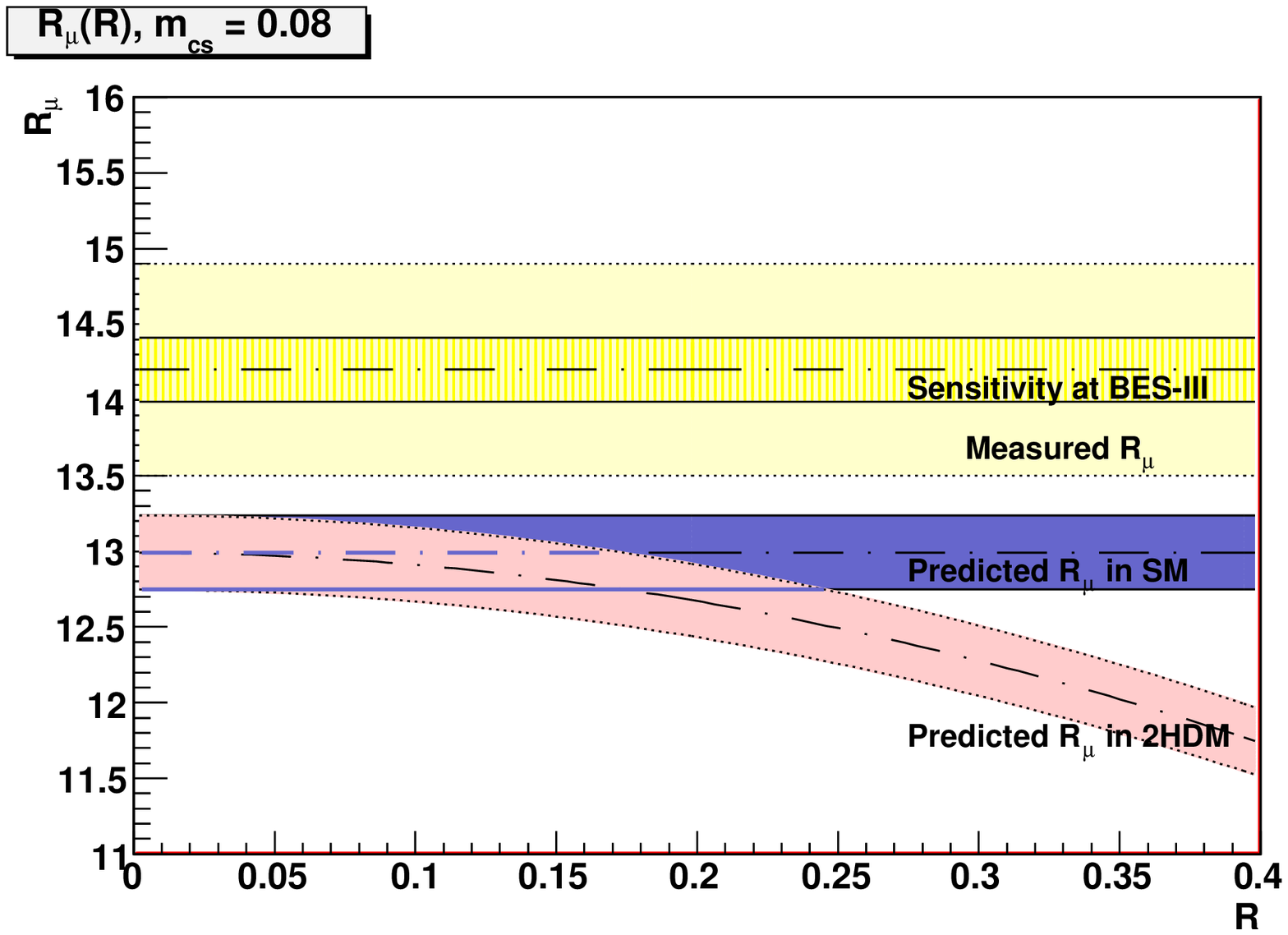}
\figcaption{\label{fig:exp:rmur} ${\cal R}_{\mu}$ as a function of
$R = \mbox{tan}\beta/m_{H^\pm}$
 for $m_{sc} = m_s/(m_s + m_c)=0.08$ and $f_{\Ds}/f_{\Dp}=
  1.164 \pm 0.011$ from LQCD calculations.  The uncertainty on the
  theoretical prediction of ${\cal R}_{\mu}$ is shown as the gray band, and a
detailed discussion can be found in Ref.~\citep{akeroyd2007}.
  The expected $\pm 1\sigma$ BES-III uncertainty experimental range of
 ${\cal  R}_{\mu}$ is indiated by the yellow band.
  The sensitivity for the measurement of the
  ratio ${\cal R}$ at BES-III is about 2.6\% level
  is also shown with the assumption that the current central value for
  ${\cal R}_{\mu}$ persists.}
\end{center}

Another, more conservative approach, is to use the LQCD prediction
for the ratio $f_{\Ds}/f_{\Dp}$, which is inherently more precise
than those for the individual $f_D$ values. A significant
deviation of this ratio from the SM prediction would be a very
robust sign of new physics beyond the SM.

Experimentally, the ratio $f_{\Ds}/f_{\Dp}$ can be extracted from
the measured ratio ${\cal R}_{\mu}$ of the leptonic decay rates of
the $\Ds$ and the $\Dp$. In the SM, one has~\cite{akeroyd2007}:
\begin{eqnarray}
{\cal R}_{\mu} &\equiv& \frac{{\cal BR}(\Ds \to \mu^+ \nu)}{{\cal
BR}(\Dp  \to \mu^+ \nu)}= \left|\frac{f_{\Ds}}{f_{\Dp}}\right|^2
\left|\frac{V_{cs}}{V_{cd}}\right|^2\frac{m_{\Ds}}{m_{\Dp}}\times \nonumber \\
& &
\left(\frac{1-m^2_{\mu}/m^2_{\Ds}}{1-m^2_{\mu}/m^2_{\Dp}}\right)\times
\frac{\tau_{\Ds}}{\tau_{\Dp}}.
 \label{eq:ratio:rate}
\end{eqnarray}
In the case of the 2HDM, new physics only modifies the decay of
$\Ds$, and the ratio ${\cal R}_{\mu}$ in Eq.~\ref{eq:ratio:rate}
is corrected by a factor $r_s$ defined in Eq.~\ref{eq:define}.

Using only CLEOc measurements and the SM relation, the
experimental value  for the $f_{\Ds}/f_{\Dp}$ ratio
is~\cite{rosner2008}
\begin{eqnarray}
r_{\Ds/\Dp} \equiv \frac{f_{\Ds}}{f_{\Dp}} = 1.23 \pm 0.10,
 \label{eq:ratio_exp_cleoc}
\end{eqnarray}
The most precise prediction of the ratio from
LQCD~\cite{hpqcd2008} is $f_{\Ds}/f_{\Dp} = 1.164 \pm 0.011$,
which has a claimed precision that is better than 1\%, and an
order of magnitude better than the existing experimental
determination. The discrepancy is about 1.0 $\sigma$ between the
current experimental determination and the LQCD calculations.

In Fig~\ref{fig:exp:rmur}, ${\cal R}_{\mu}$ is plotted as a
function of $R=\mbox{tan}\beta/m_{H^\pm}$ for the case of the
2HDM, using $m_{sc} = m_s/(m_s + m_c)=0.08$ and $f_{\Ds}/f_{\Dp}=
1.164 \pm 0.011$ from the LQCD calculation (detailed discussion on
2HDM in Ref.~\cite{akeroyd2007}).   The SM prediction for ${\cal
R}_{\mu}$ is $(12.99 \pm 0.25)$, where the error is from the
uncertainty on the LCQCD prediction for $f_{\Ds}/f_{\Dp}$.
Compared to the measured value ${\cal R}_{\mu} = 14.2 \pm 0.7$, we
see that the SM prediction is almost 2 standard deviations lower.
If the LQCD calculation is reliable, this indicates that we need a
modification to the SM that has constructive interference to
accommodate the discrepancy~\cite{bogdan2008}. It may be concluded
that the 2HDM discussed in Ref.~\cite{akeroyd2007} is disfavored
by the current data. It would be very interesting if the
experimental precision on ${\cal R}_{\mu}$ ratio could be improved
to match the one percent level of the theoretical errors in the
near future. As discussed below in table~\ref{tab:constants:bes3},
the sensitivity of the measurement of the ratio at BES-III is
about 2.6\% with 20 fb$^{-1}$ at $E_{CM} = 4170$
MeV~\cite{li2006}. This results in a 1.0\% uncertainty on the
ratio of $f_{\Ds}/f_{\Dp}$.

\begin{center}
\tabcaption{\label{tab:lum} $\tau$-Charm productions at BEPC-II in
one year's running ($10^7s$)~\citep{li2006}.}
\begin{tabular}{@{}lll}
\hline
               & Central-of-Mass    & \#Events  \\
Data Sample    & (MeV)                   & per year  \\
\hline
$J/\psi$ &  3097     & $10\times 10^9$\\
$\tau^+\tau^-$   & 3670  & $12\times 10^6$ \\
$\psi(2S)$ & 3686  & $3.0\times 10^9$ \\
$\DzDzb$ & 3770  & $18\times 10^6$ \\
$\DpDm$ & 3770  & $14\times 10^6$ \\
$\DspDsm$ & 4030  & $1.0\times 10^6$ \\
$\DspDsm$ & 4170  & $2.0\times 10^6$ \\
\hline
\end{tabular}
\end{center}

Beginning in mid-2008, the BEPC-II/BES-III will be operated at
center-of-mass (CM) energies corresponding to $\sqrt{s} = 2.0 -4.6
$ GeV. The designed luminosity over this energy region will range
from $1\times 10^{33}$cm$^{-2}$s$^{-1}$ down to about $0.6 \times
10^{33}$cm$^{-2}$s$^{-1}$~\cite{bepcii}, yielding around 5
fb$^{-1}$ each at $\psi(3770)$ and at $\sqrt{s} = 4170$
MeV~\cite{bepcii} above $\DspDsm$ threshold and 3 fb$^{-1}$ at
$J/\psi$ peak in one year's running with full
luminosity~\cite{bepcii}. These integrated luminosities correspond
to samples of 2.0 million $\DspDsm$, 30 million $\DD$ pairs and
$10\times 10^9$ $J/\psi$ decays. Table~\ref{tab:lum} summarizes
the data set per year at BES-III. In this paper,  the sensitivity
studies are based on 20 fb$^{-1}$ luminosity at $\psi(3770)$ peak
for $D$ physics, the same luminosity also for $D_s$ physics at
$\sqrt{s} = 4170$ MeV.

According to the recent energy scan above the threshold of $D^+_s
D^-_s$ pair from CLEO-c~\cite{newscan2006}, the production cross
section of $D^{*+}_s D^-_s + D^+_s D^{*-}_s$ is about 1.0 nb at
4170 MeV, which is 3 times higher than the cross section of $D^+_s
D^+_s$ at 4030 MeV. The scan succeeded in identifying a $\Ds$
¡°minifactory¡± at $E_{CM}=4170$ MeV.  At this energy the $e^+
e^-$ annihilation cross sections into $D^{(*)}_{(s)}$ pairs are
estimated from the scan of CLEO-c and summarized in
table~\ref{tab:cross}.

\begin{center}
\tabcaption{\label{tab:cross}The preliminary results for the cross
sections of the $D^{(*)}_{(s)}$ pairs at $E_{CM}=4016$ MeV and
$E_{CM}=4170$ MeV, respectively,  from CLEO-c
experiment~\citep{newscan2006}.}
\begin{tabular}{@{}lll}
\hline
               & 4016 MeV     &  4170 MeV  \\
                &  $D^*\bar{D}^*$ threshold  &    \\
Decay Modes     &  in nb &  in nb \\

\hline
$\sigma(D^+_s D^-_s)$      & 0.25       & $<0.05$     \\
$\sigma(D^{*+}_s D^-_s + D^+_s D^{*-}_s)$ &  - & 1.0 \\
$\sigma(D \bar{D}^{*})+D^{*} \bar{D})$ & 7.0  & 2.0 \\
$\sigma(D^{*} \bar{D}^{*})$ & 3.0     & 5.1 \\\hline
\end{tabular}
\end{center}

\begin{center}
\tabcaption{\label{tab:constants:bes3} Expected errors on the
branching fractions for leptonic decays and decay constants at the
BES-III with 20 fb$^{-1}$ at $\psi(3770)$ peak and $E_{CM}= 4170$
MeV, respectively.}
\begin{tabular}{c|c||c|c}
\hline
Observable & Error   & Measurement & Error \\
\hline
${\cal BR}(\Dp  \to \mu^+ \nu)$& 2.0\%     & $f_D|V_{cd}|$  & 1.1\%  \\
${\cal BR}(\Ds \to \mu^+ \nu)$ &  2.0\%    & $f_{Ds}|V_{cs}|$ & 1.0\%  \\
$\frac{{\cal BR}(\Ds \to \mu^+ \nu)}{{\cal BR}(\Dp  \to \mu^+
\nu)}$  & 2.6\% &
$\left|\frac{V_{cs}f_{Ds}}{V_{cd}f_D}\right|$ & 1.3\% \\
\hline
\end{tabular}
\end{center}

\section{Decay constants at BES-III}
Measurements of leptonic decays at the BES-III will benefit from
the fully tagged $D^+$ and $D^-_s$ decays available at the
$\psi(3770)$ and at $\sqrt{s} \sim 4170$ MeV.
 The leptonic decay of $D^+_s (D^+) \rightarrow \mu^+
\nu$ is detected by using this kind of "double-tag" techniques in
which one $\Dp$ or $\Ds$ is fully reconstructed and the rest of
the event is examined without bias but with substantial kinematic
constraints~\cite{asnercharm2007}.   For the decay of $\Dp
\rightarrow \mu^+\nu_{\mu}$ at $\psi(3770)$ peak, the pure $\DD$
pair in the initial state and cleanliness of the full tag
reconstruction make this measurement essentially background-free
at CLEO-c and BES-III~\cite{cleocfd,li2006}. The leptonic decay
rate for $D^+$ can be measured with a precision of 1-2\% level at
the BES-III experiment. This will allow the validation of
theoretical calculations of the decay constants at the 1\% level.
Table~\ref{tab:constants:bes3} summarizes the expected precision
in the decay constant measurements.

For the decay of $\Ds \rightarrow \mu \nu_{\mu}$ at $E_{CM}= 4170
$ MeV,  to select the sample of single tag events, one has to
fully reconstruct one of $D_s$ by using the decay modes, such as
$\Ds \rightarrow K^+K^-\pi^+$, $K_sK^+$, $\eta (\eta^{\prime})
\pi^+$, $\pi^+\pi^-\pi^+$ and $K^{*+}K^{*0}$ as described in
reference~\cite{prdcleoc2007} in the CLEO-c experiment.  Then one
can find another photon to reconstruct the $D^*_s \rightarrow D_s
\gamma$ decay. For the $D^*_s D_s$ candidates,  the missing
mass-squared, $MM^{*2}$, recoiling against the photon and the
$D_s$ tag should peak at the $D_s$ mass-squared, one obtains:
\begin{eqnarray}
MM^{*2}= (E_{CM}-E_{D_s} -E_{\gamma})^2 - (\vec{p}_{CM} -
\vec{p}_{D_s} - \vec{p}_{\gamma})^2, \nonumber
 \label{eq:missingmass}
\end{eqnarray}
where $E_{CM}$ ($\vec{p}_{CM}$) is the center-of-mass energy
(momentum), $E_{D_s}$ ($\vec{p}_{D_s}$) is the energy (momentum)
of the fully reconstructed $\Ds$ tag, and $E_{\gamma}$
($\vec{p}_{\gamma}$) is the energy (momentum) of the additional
photon.

Candidates $\Ds \rightarrow \mu^+ \nu$ events are reconstructed by
selecting events with only a single extra muon with opposite sign
of charge to the tag side. Thus, the undetected energy and
momentum is interpreted as the neutrino four-vector,  the missing
mass squared, $MM^2$, evaluated by taking into account the seen
muon, $\Ds$, and the photon should peak at zero, and is given by
\begin{eqnarray}
MM^{2}&=& (E_{CM}-E_{D_s} -E_{\gamma}-E_{\mu})^2 -\nonumber \\
&&(\vec{p}_{CM} -  \vec{p}_{D_s} -
\vec{p}_{\gamma}-\vec{p}_{\mu})^2,
 \label{eq:missnutrino}
\end{eqnarray}
where $E_{\mu}$ ($\vec{p}_{\mu}$) is the energy (momentum) of the
candidate muon. The missing mass resolution is about one pion
mass~\cite{asnercharm2007}. These techniques pioneered by Mark-III
and well developed by the CLEO-c~\cite{heq2006} allow precise
absolute branching fraction determination. Backgrounds can be
highly suppressed and the statistical errors and systematic errors
can be minimized~\cite{asnercharm2007}.

 At BES-III, in order to measure the branching fractions of the
pure leptonic decays at 2\% level or below, one has to pay more
attention to the systematic uncertainties due to the dilution to
the signal region from irreducible backgrounds. The main
backgrounds are from $\Ds \rightarrow \pi^+ \pi^0$, $\Ds
\rightarrow \pi^+ \eta$, and $\Ds \rightarrow \pi^+ K_L$ in which
the neutral particles in the final states are not detected and
only a charged pion left.  To suppress the background with only a
detected charged pion, at BES-III,  both calorimeter and muon
counters can be used to distinguish the muon from the pion. Since
the typical momentum of muon and pion from $\Ds$ decays are above
700 MeV,  according to the design of BES-III~\cite{bepcii}, the
detection efficiency of muon in the muon counters is above 97\%,
and the rate of contamination from pion is less than 5\%. Thus, if
we take the estimated 90\% upper limit for the $\Ds \rightarrow
\pi^+ \pi^0$ decay as $1.1 \times 10^{-3}$, and find zero expected
background under the signal peak. Another dangerous background is
from the radiative decay $\Ds \rightarrow \mu^+ \nu_{\mu} \gamma$.
For $\Ds \rightarrow \mu^+ \nu$ radiative corrections had been
estimated and found that it is genuinely of order
$\alpha$~\cite{burdman1995}.  In the analysis of
CLEO-c~\cite{prdcleoc2007}, a cut $E_{\gamma}<300 MeV$ is used to
remove the dilution from radiative decay,  and one finds that the
radiative rate is less than 1\% which can be neglected with
current statistics.

\section{Summary}
In summary, following the work of Rosner~\cite{rosner2008} and
Dobrescu {\it et al.}~\cite{bogdan2008}, we have reviewed the most
recent determinations of the decay constant $f_{\Ds}$ from
different experiments.  We find that the current experimental
determination in the SM differs from the most precise unquenched
lattice QCD calculation at the 4 $\sigma$ level. Meanwhile, the
measured ratio, ${\cal BR}(\Ds \rightarrow \mu^+
\nu_{\mu})$/${\cal BR}(\Dp \rightarrow \mu^+ \nu_{\mu})$, is
larger than the standard model prediction at the 2.0$\sigma$
level. With current data, the occurrence of new physics, in the
case of the 2HDM~\cite{akeroyd2007}, is disfavored.  The measured
ratio, ${\cal BR}(\Ds \rightarrow \mu^+ \nu_{\mu})$/${\cal BR}(\Dp
\rightarrow \mu^+ \nu_{\mu})$, suggests that we need a new physics
contribution with constructive interference. we discuss that the
precise measurement of the ratio ${\cal BR}(\Ds \rightarrow \mu^+
\nu_{\mu})$/${\cal BR}(\Dp \rightarrow \mu^+ \nu_{\mu})$ at
BES-III will shed light on the presence of new intermediate
particles by comparing with the theoretical predictions,
especially, the predictions of high precise unquenched lattice QCD
calculation.

{\bf Acknowledgements} H.~B.~Li would thank D.~M.~Asner,
A.~G.~Akeroyd, S. Olsen and M.~Z.~Yang for useful discussions and
suggestions. This work is supported in part by the National
Natural Science Foundation of China under contracts Nos.
10575108,10521003, 10735080, the 100 Talents program of CAS, and
the Knowledge Innovation Project of CAS under contract Nos. U-612
and U-530 (IHEP).

\end{multicols}

\vspace{-2mm} \centerline{\rule{80mm}{0.1pt}} \vspace{2mm}

\begin{multicols}{2}


\end{multicols}

\end{document}